\documentstyle[aps,prb,epsf,floats]{revtex}

\begin{document}
\draft

\renewcommand{\floatpagefraction}{0.8}

%************************************************************
%***** Title ************************************************
%************************************************************

\twocolumn[\hsize\textwidth\columnwidth\hsize\csname@twocolumnfalse%
\endcsname
\title{Orbital polarons in the metal-insulator transition of manganites}
\author{R.\ Kilian}
\address{Max-Planck-Institut f\"ur Physik komplexer Systeme,
N\"othnitzer Strasse 38, D-01187 Dresden, Germany}
\author{G.\ Khaliullin}
\address{Max-Planck-Institut f\"ur Festk\"orperforschung,
Heisenbergstrasse 1, D-70569 Stuttgart, Germany}
\date{\today}
\maketitle

%************************************************************
%***** Abstract *********************************************
%************************************************************

\begin{abstract}
The metal-insulator transition in manganites is strongly influenced
by the concentration of holes present in the system. Based upon an
orbitally degenerate Mott-Hubbard model we analyze two possible 
localization scenarios to account for this doping dependence:
First, we rule out that the transition is initiated by a disorder-order
crossover in the orbital sector, showing that its effect on charge 
itineracy is only small. Second, we introduce the idea of orbital polarons 
originating from a strong polarization of orbitals in the vicinity of holes. 
Considering this direct coupling between charge and orbital
degree of freedom in addition to lattice effects we are
able to explain well the phase diagram of manganites for
low and intermediate hole concentrations.
\end{abstract}

\rule{0mm}{8mm}]

%**********************************************************
%*** Introduction *****************************************
%**********************************************************

\section{Introduction}

The doping dependence of the properties of manganese oxides poses some 
of the most interesting open problems in the physics of these compounds.
First to be noticed is the peculiar asymmetry of the phase diagram that 
is most pronounced in the charge sector:
Regions of high ($x>0.5$) and low ($x<0.5$) concentration of holes are
characterized by such contrasting phenomena as charge ordering and 
metalicity, respectively. In the latter region~--- which we wish to focus
on~--- the metallic state can be turned into an insulating one by  
raising the temperature above the Curie temperature $T_C$.
Introducing the notion of double exchange
which associates the relative orientation of localized Mn $t_{2g}$ 
spins with the mobility of itinerant $e_g$ electrons,
early work has identified this transition to be controlled by the 
loss of ferromagnetic order inherent to the metallic state.\cite{ZAD}
It is believed that lattice effects are also of crucial 
importance in this transition. Within the lattice-polaron 
double-exchange picture,\cite{MIL,ROE96} the crossover from metallic to 
insulating behavior is controlled by the ratio of  polaron binding 
energy $E_b$ to the kinetic energy $E_{\text{kin}}$ of charge carriers:
\begin{equation}
\lambda = \frac{E_b}{E_{\text{kin}}}.
\label{LCR}
\end{equation}
When forming a bound state with the lattice, charge carriers loose part 
of their kinetic energy. Hence, polarons are stable only if 
this loss in energy is more than compensated by the gain in binding 
energy, i.e., if $\lambda > 1$. In a double-exchange system, this
critical coupling strength may be reached by raising the 
temperature~--- the double-exchange mechanism then acts to 
reduce the kinetic energy and hence to increase $\lambda$.
Spin disorder and spin-polaron effects further enhance
the carrier localization above $T_C$.\cite{VAR96} The doping dependence 
of the metal-insulator transition, however, is not fully captured 
in this picture. Namely the complete breakdown 
of metalicity at hole concentrations below $x_{\text{crit}} \approx 
0.15$ - $0.2$ that occurs despite the fact that ferromagnetism is
fully sustained remains an open problem which we address in this paper.

The effective coupling constant $\lambda$ in Eq.\ (\ref{LCR})
has originally been introduced for non-interacting electrons. 
The itinerant $e_g$ electrons in manganites, on the other hand, are 
subject to strong on-site repulsions which necessitates to accommodate
the definition of $\lambda$. According to numerical studies,
\cite{CAP} the basic physics underlying Eq.\ (\ref{LCR}) 
remains valid even in correlated systems: As in the free-electron 
case, the metal-insulator transition is controlled by the competition
between the polaron binding energy and the kinetic energy of charge carriers. 
Nevertheless,
correlation effects might renormalize these two relevant energy scales,
presumably introducing a doping dependence.
In fact, the Gutzwiller bandwidth of correlated electrons scales
with the concentration of doped holes; one could therefore
be inclined to set $E_{\text{kin}} \propto xt$, where $t$ denotes the 
hopping amplitude. But this approach reaches too short:
The Gutzwiller picture describes only the average kinetic energy of
the system. In contrast, the relevant quantity for localizing the
holes doped into a Mott-Hubbard system is the characteristic energy 
scale of charge fluctuations which remains $\propto t$.\cite{DAG94}
Pictorially this quantity corresponds to the kinetic energy of
a single hole. We thus conclude that a more  
thorough treatment of correlation effects is needed in order to explain the 
peculiar doping dependence of the metal-insulator transition in manganites.

In this paper we analyze two mechanisms that could drive the localization 
of charge carriers at small hole concentrations $x$. First we explore the
possibility of the metal-insulator transition to be controlled by a 
disorder-order crossover in the $e_g$-orbital sector. The idea is the 
following: Orbital fluctuations are induced by the motion of holes 
and hence possess an energy scale $\propto xt$. At large $x$,
orbitals fluctuate strongly and inter-site orbital correlations are
weak. As the concentration of holes is reduced, fluctuations slow 
down until a critical value of $x$ has been reached~--- promoted by
Jahn-Teller and superexchange coupling, an orbital-lattice 
ordered state now evolves. We analyze the extent to which this 
transition in the orbital-lattice sector affects the itineracy of holes.
Finding almost similar values for the kinetic energy of doped holes
in orbitally ordered 
and disordered states we are lead to conclude that the development of 
orbital-lattice order is not sufficient to trigger the localization process. 
Next we turn to analyze a second scenario of the metal-insulator transition 
for which we introduce the concept of orbital polarons.
Similar to spin polarons in correlated spin systems, 
orbital polarons are a natural consequence of strong electron correlations 
and the double degeneracy of on-site levels~--- in manganites the latter 
follows from the degeneracy of $e_g$ orbitals. We argue that holes polarize 
the orbital state of $e_g$ electrons on neighboring sites:
A splitting of orbital levels is evoked by a displacement of oxygen ions 
and also by the Coulomb force exerted by the positively charged holes. Being
comparable in magnitude to the kinetic energy of holes, the orbital-hole 
binding energy can be large enough for holes and surrounding orbitals to 
form a bound state. The important point is that the stability of these orbital
polarons competes not only against the kinetic energy of holes but also
against the fluctuation rate $\propto xt$ of orbitals: The faster the latter 
fluctuate, the less favorable it is to form a bound state in which orbitals
have to give up part of their fluctuation energy. Combining this new 
orbital-polaron
picture with that of conventional lattice polarons we are able to explain well 
the phase diagram of manganites at low and intermediate doping levels.

%**********************************************************
%*** Orbital Disorder-Order Transition *******************
%**********************************************************

\section{Orbital Disorder-Order Transition}
\label{SEC:ORB}

In this section we analyze the impact of a disorder-order transition
in the orbital-lattice sector onto the itineracy of holes.
Our motivation is that a sudden freezing out of 
orbital fluctuations below a critical doping concentration could 
significantly impede the motion of holes, hence initiating
the metal-insulator transition. By comparing the bandwidth of holes 
both in orbitally ordered and disordered states, we are able
to refute this idea: The orbital sector is shown to have only little 
influence onto the charge mobility in manganites.

%**********************************************************

\subsection{Disordered State}

We begin by investigating the bandwidth of holes in a strongly
fluctuating, orbitally disordered state.
Our starting point is the $t$-$J$ model of double-degenerate
$e_g$ electrons which, via Hund's coupling, interact 
ferromagnetically with an array of localized $S=3/2$ core spins.
The model accounts for the presence of strong on-site repulsions 
that forbids more than one $e_g$ electron to occupy the same Mn 
site as well as for the double-degeneracy of $e_g$ levels.
At low temperatures and intermediate doping levels, the 
double-exchange mechanism induces a
parallel alignment of spins. Treating deviations from this
ferromagnetic ground state only on a mean-field level as is discussed 
below, the core spins can be discarded and the spin indices of $e_g$ 
electrons may be dropped; the $t$-$J$ Hamiltonian 
then becomes\cite{ISH97b,KIL98}
\begin{equation}
H_{tJ} = - \sum_{\langle ij\rangle_{\gamma}} \left(
t^{\alpha\beta}_{\gamma} \hat{c}^{\dagger}_{i\alpha}
\hat{c}_{j\beta} + \text{H.c.}\right)
+\frac{2J}{z}\sum_{\langle ij \rangle_{\gamma}} \tau^{\gamma}_i \tau^{\gamma}_j,
\label{TRH}
\end{equation}
with $z=6$. Nearest-neighbor bonds along spatial directions 
$\gamma\in\{x,y,z\}$ are denoted by $\langle ij \rangle_{\gamma}$.
We use constrained operators 
$\hat{c}^{\dagger}_{i\alpha} = c^{\dagger}_{i\alpha}\left(1-n_i\right)$
which create an $e_g$ electron at site $i$ in orbital $\alpha$ only
under the condition that the site is empty. The first term in
Eq.\ (\ref{TRH}) describes the inter-site transfer of constrained $e_g$ 
electrons. The transfer amplitude depends upon the orientation
of orbitals at a given bond as is reflected by the transfer 
matrices
\[
t_{x/y}^{\alpha\beta} = 
t\left(\begin{array}{cc}
1/4 & \mp\sqrt{3}/4\\
\mp\sqrt{3}/4 & 3/4
\end{array}\right), \quad
t_{z}^{\alpha\beta} = 
t\left(\begin{array}{cc}
1 & 0\\
0 & 0
\end{array}\right);
\]
a representation with respect to the orbital basis 
$\alpha \in \{|3z^2-r^2\rangle, |x^2-y^2\rangle \}$ has been chosen 
here. Due to its non-diagonal structure, orbital quantum numbers are 
not conserved by Hamiltonian (\ref{TRH})~-- inter-site transfer processes 
induce fluctuations in the orbital sector. 
The second term in Eq.\ (\ref{TRH}) accounts for processes involving
the virtual occupation of sites by two $e_g$ electrons. This superexchange 
mechanism establishes an inter-site coupling between orbital pseudospins of 
overall strength $J=zt^2/U$, where $U$ is the on-site repulsion between 
spin-parallel
$e_g$ electrons. The pseudospin operators are
\begin{equation}
\tau^{x/y}_i = -\frac{1}{4}(\sigma^z\pm\sqrt{3}\sigma^x), \quad
\tau^z_i = \frac{1}{2}\sigma^z,
\label{PSE}
\end{equation}
with Pauli matrices $\sigma^{x/z}_i$ acting on the orbital subspace.
Jahn-Teller phonons mediate an additional interaction 
between orbital pseudospins which is of the exact same form as the
superexchange term. The numerical value of $J$ has to be chosen 
such as to comprise both effects. We finally note that deviations
from the ferromagnetic ground state underlying Hamiltonian (\ref{TRH})
are treated within conventional double-exchange theory. The transfer 
amplitude $t$ then depends on the normalized magnetization $m$ 
via\cite{MIL,KUB72}
\begin{equation}
t=t_0\sqrt{(1+m^2)/2},
\label{DOX}
\end{equation}
where $t_0$ denotes the hopping amplitude between spin-parallel
Mn sites.

To observe the strongly correlated nature of $e_g$ electrons,
it is convenient to introduce separate particles for charge
and orbital degrees of freedom.\cite{ISH97a} The metallic phase of 
manganites can be well described within an orbital-liquid picture
that accounts for orbital fluctuations by employing a slave-boson 
representation of electron operators:\cite{KIL98,KHA99}
\[
c^{\dagger}_{i\alpha} = f^{\dagger}_{i\alpha}b_i.
\]
Here orbital pseudospin is carried by fermionic orbitons 
$f_{i\alpha}$ and charge by bosonic holons $b_i$.
Introducing mean-field parameters
$\chi=t^{-1}\sum_{\alpha\beta} t^{\alpha\beta}_{\gamma} \langle
f^{\dagger}_{i\alpha} f_{i\beta}\rangle$ and
$x=\langle b^{\dagger}_i b_j\rangle$,
where $x$ is the concentration of holes and $\chi\approx \frac{1}{2}$, 
the two types of quasiparticles can be decoupled:
\begin{eqnarray}
H_{\text{orb}} &=& -\left(x+\frac{2\chi J}{zt}\right)
 \sum_{\langle ij\rangle_{\gamma}}
t^{\alpha\beta}_{\gamma} \left(f^{\dagger}_{i\alpha} f_{j\beta}
+\text{H.c.}\right),
\label{OHA}\\*
H_{\text{hl}} &=& -\chi t \sum_{\langle ij \rangle} 
\left( b^{\dagger}_i b_j + \text{H.c.}\right).
\label{CAH}
\end{eqnarray}
Diagonalizing the above expressions in the momentum representation one
obtains
\[
H_{\text{orb}} = \sum_{\bbox{k}\nu}\xi^{\nu}_{\bbox{k}}
f^{\dagger}_{\bbox{k}\nu} f_{\bbox{k}\nu},\quad
H_{\text{hl}} = \sum_{\bbox{k}} \omega_{\bbox{k}}
b^{\dagger}_{\bbox{k}}b_{\bbox{k}},
\]
with index $\nu=\pm$ and dispersion functions
\begin{eqnarray}
\xi^{\pm}_{\bbox{k}} &=& (xt+\frac{2\chi J}{z})
\Big[-\epsilon_0(\bbox{k})\pm\sqrt{\epsilon_1^2(\bbox{k}) +
\epsilon_2^2(\bbox{k})}\Big],\nonumber\\*
\omega_{\bbox{k}} &=& 6\chi t\Big[1-\frac{1}{3}
\epsilon_0(\bbox{k})\Big],\nonumber
\end{eqnarray}
where
$\epsilon_0(\bbox{k}) = c_x+c_y+c_z$,
$\epsilon_1(\bbox{k}) = (c_x+c_y)/2-c_z$,
$\epsilon_2(\bbox{k}) = \sqrt{3}(c_x-c_y)/2$
with $c_{\gamma} = \cos k_{\gamma}$.
The essence of this slaved-particle mean-field treatment is that orbital 
and charge fluctuations are assigned different energy scales. This
is reflected by the bandwidths of orbiton and holon quasiparticles, 
respectively:
\begin{eqnarray}
W_{\text{orb}} &=& 6xt+J,
\label{OFL}\\*
W_{\text{hl}} &=& 6t.
\label{CFL}
\end{eqnarray}
The former quantity $W_{\text{orb}}$ sets the energy
scale of orbital fluctuations~--- the terms proportional to
$xt$ and $J$ describe fluctuations induced by the motion 
of holes and by the coupling between pseudospins, respectively. The 
latter quantity $W_{\text{hl}}$ finally defines the itineracy 
of holes in the orbital-liquid state. The variation of
the holon bandwidth with the onset of orbital order is in the
focus of our interest in the remainder of this section.

%**********************************************************

\subsection{Instability Toward Orbital Order}

The above treatment of orbital and charge fluctuations is based
upon the notion of a strongly fluctuating orbital state that is far 
from any instability towards orbital order. In real systems
such instabilities do exist: Jahn-Teller phonons and
superexchange processes mediate a coupling between orbitals
on neighboring sites which introduces a bias towards orbital-lattice
ordering. Competing 
against the energy scale of orbital fluctuations $\propto xt$, order 
in the orbital sector is expected to evolve below a critical doping 
concentration $x_{\text{crit}}$. We investigate this instability of 
the orbital-liquid state by introducing the inter-site coupling term
\begin{equation}
H_J = -\frac{2J}{z} \sum_{\langle ij\rangle_{\gamma}}
\tau_i^{\Theta} \tau_j^{\Theta} \, e^{iQ_{\gamma}},
\label{HCP}
\end{equation}
with $z=6$ and $\tau_i=(\sin\Theta\sigma^x_i+\cos\Theta 
\sigma^z_i)/2$ acting on the orbital subspace. We note that 
Eq.\ (\ref{HCP}) is a simplification of the superexchange coupling 
term in Hamiltonian (\ref{TRH})~--- internal frustration makes the 
latter difficult to handle. For $\Theta=\pi/2$ and
$\bbox{Q}=(\pi,\pi,0)$, the pseudospin interaction
in Eq.\ (\ref{HCP}) favors a staggered-type orbital orientation
\begin{equation}
|\Theta\rangle^{\pm} = \left(|3z^2-r^2\rangle \pm |x^2-y^2\rangle\right)/
\sqrt{2}
\label{UPS}
\end{equation}
within $x$-$y$ planes repeating itself along the $z$ direction; 
this closely resembles the type of order observed experimentally in 
LaMnO$_3$.\cite{MUR98}
The breakdown of the orbitally disordered state, i.e., the development of 
orbital order, is signaled by a singularity in the static orbital
susceptibility $\langle \sigma_{\bbox{Q}}^x\sigma_{-\bbox{Q}}^x
\rangle_{\omega=0}$. Employing a random-phase approximation, the 
latter can be expressed as
\begin{equation}
\langle \sigma^x\sigma^x\rangle_{\bbox{Q}} = 
\frac{\langle\sigma^x\sigma^x\rangle^0_{\bbox{Q}}}
{1+J_{\bbox{Q}}\langle\sigma^x\sigma^x\rangle^0_{\bbox{Q}}/2},
\label{IRP}
\end{equation}
with vertex function $J_{\bbox{q}} = J \left(\cos q_x + \cos q_y - \cos q_z
\right)/3$ and the shorthand notation $\langle \sigma^x\sigma^x
\rangle_{\bbox{Q}}=\langle \sigma_{\bbox{Q}}^x\sigma_{-\bbox{Q}}^x 
\rangle_{\omega=0}$. Bare orbital 
susceptibilities $\langle \ldots \rangle^0$ are evaluated using orbiton 
propagators associated with the mean-field Hamiltonian (\ref{OHA}).
Numerically solving for the pole in Eq.\ (\ref{IRP}), we find the following 
expression for the critical doping concentration:
\begin{equation}
x_{\text{crit}} \approx \frac{J}{4t},
\label{XCR}
\end{equation}
which is valid for $x_{\text{crit}}\ll 0.5$.
At concentrations below this critical value an orbitally ordered state
is to be expected. With $J=0.13$~eV as estimated from the structural 
phase transition observed in LaMnO$_3$ at $T=780$~K 
(Ref.\ \onlinecite{MUR98}) and 
$t=0.36$~eV (Ref.\ \onlinecite{KIL98}) we obtain a critical doping 
concentration of $x_{\text{crit}} = 9\%$. 
This result indicates that the metallic state of manganites is indeed 
instable towards orbital-lattice ordering at doping concentrations
that are not too far from those at which the system is observed to 
become insulating.

%**********************************************************

\subsection{Ordered State}

\begin{figure}
\noindent
\centering
\setlength{\unitlength}{0.98\linewidth}
\begin{picture}(1,0.65)
\put(0,0.5){(a)}
\put(0,0.13){(b)}
\put(0.09,0.4){\epsfxsize=0.4\unitlength\epsffile{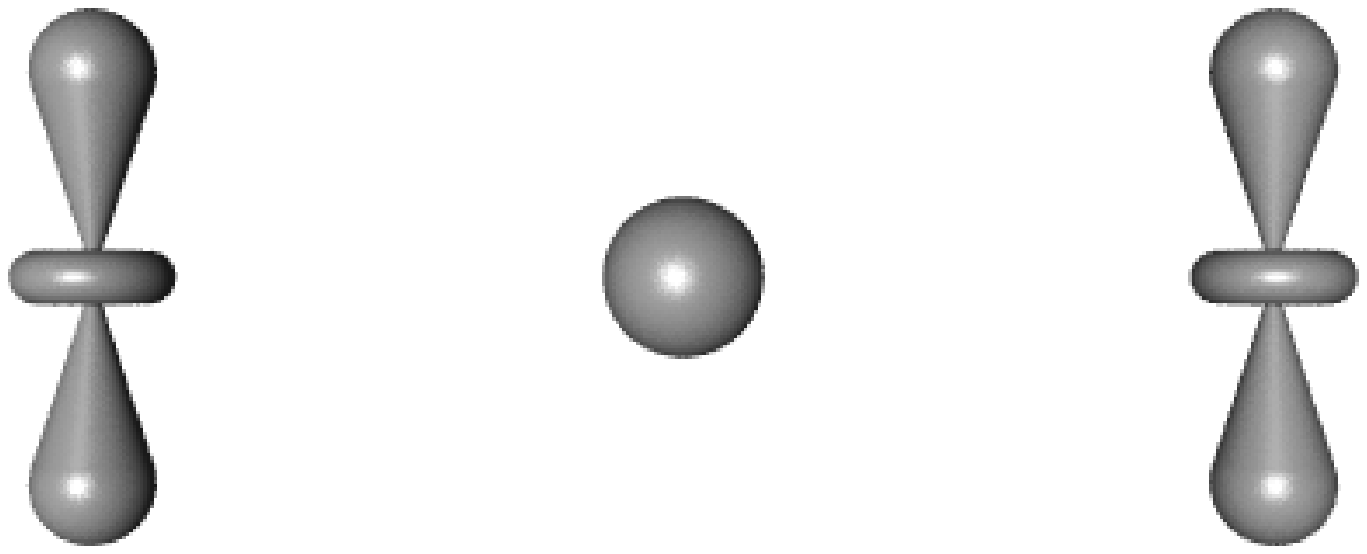}}
\put(0.56,0.4){\epsfxsize=0.4\unitlength\epsffile{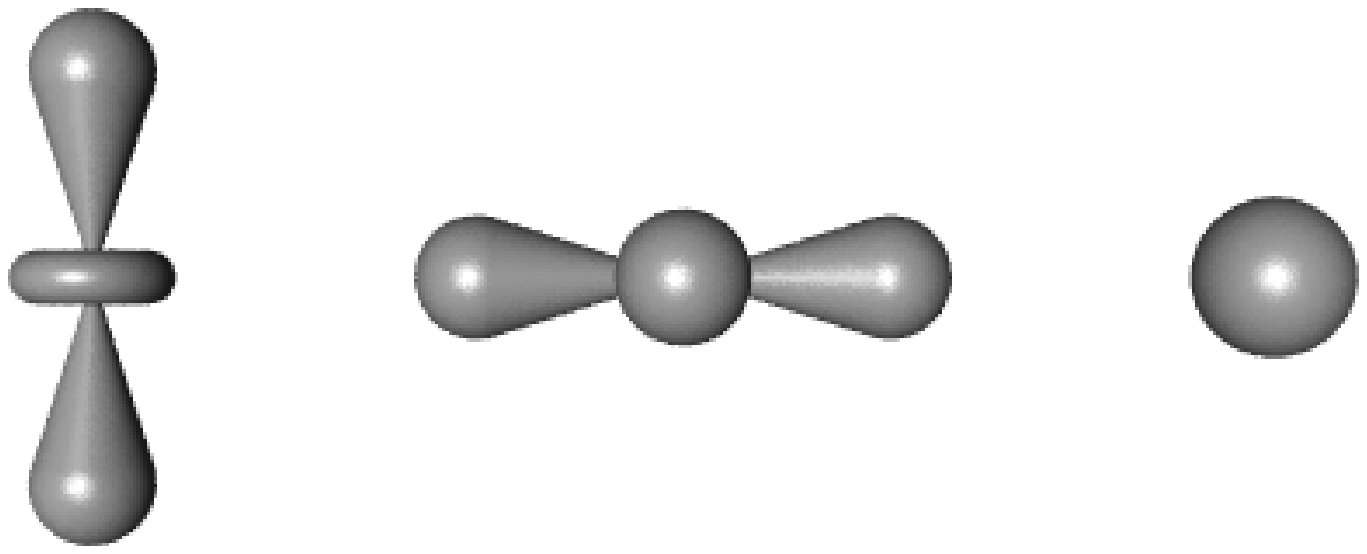}}
\put(0.09,0.1){\epsfxsize=0.4\unitlength\epsffile{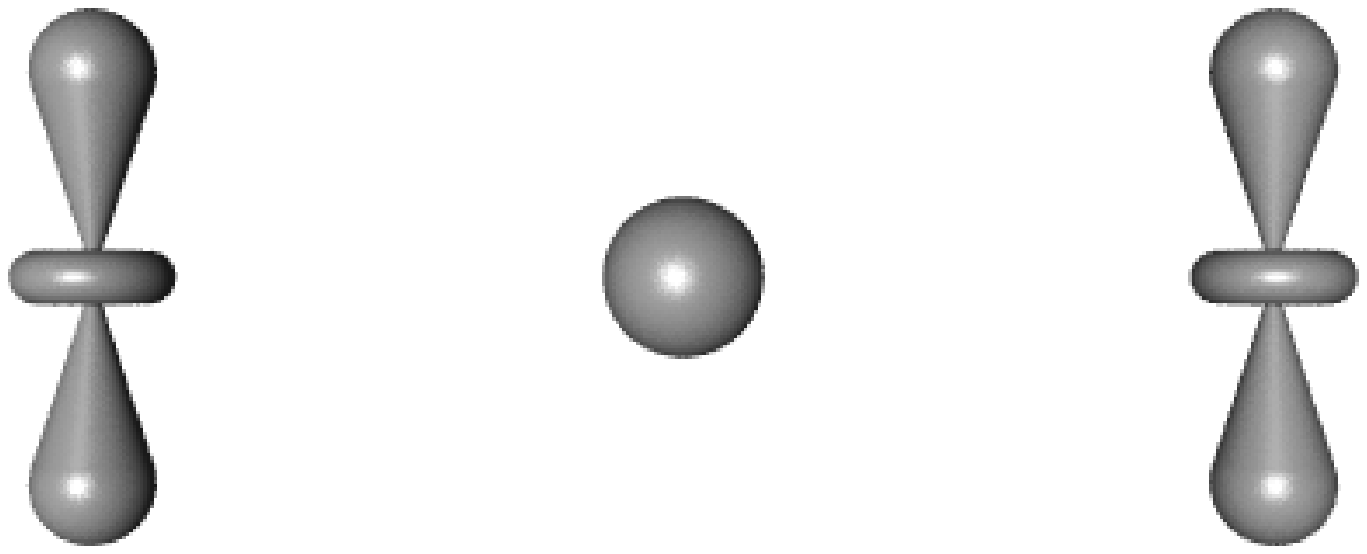}}
\put(0.56,0.1){\epsfxsize=0.4\unitlength\epsffile{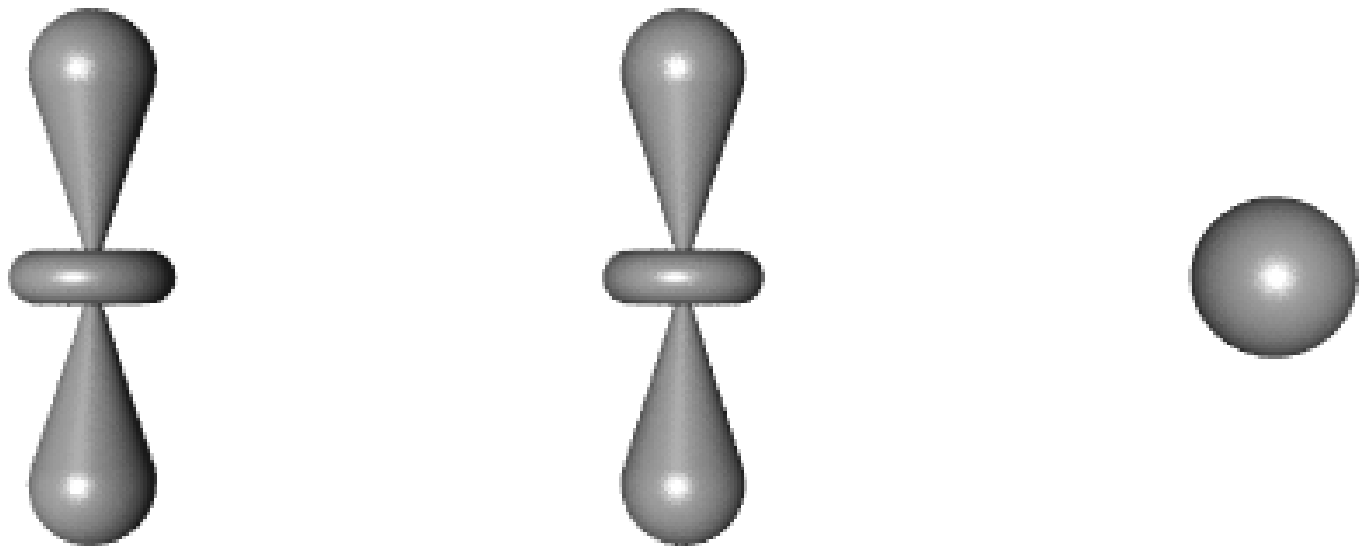}}
\put(0.08,0.40){\fbox{\rule{0.4\unitlength}{0mm}\rule{0mm}{0.2\unitlength}}}
\put(0.55,0.40){\fbox{\rule{0.4\unitlength}{0mm}\rule{0mm}{0.2\unitlength}}}
\put(0.08,0.04){\fbox{\rule{0.4\unitlength}{0mm}\rule{0mm}{0.2\unitlength}}}
\put(0.55,0.04){\fbox{\rule{0.4\unitlength}{0mm}\rule{0mm}{0.2\unitlength}}}
\put(0.24,0.305){State 1}
\put(0.7,0.305){State 2}
%\put(0.4,0.3){\epsfxsize=0.25\unitlength\epsffile{arsep.eps}}
%
\put(0.3,0.55){\epsfxsize=0.13\unitlength\epsffile{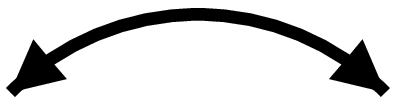}}
\put(0.3,0.05){\epsfxsize=0.13\unitlength\epsffile{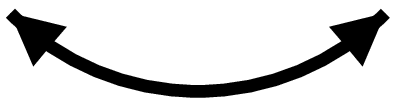}}
\put(0.34,0.52){$t^{\uparrow\downarrow}$}
\put(0.34,0.08){$t^{\uparrow\uparrow}$}
\put(0.655,0.155){\epsfxsize=0.08\unitlength\epsffile{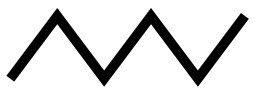}}
\put(0.675,0.115){$J$}
\end{picture}
\caption{Coherent (a) and incoherent (b) hole motion in 
antiferro-type orbital order. Incoherent processes involve
the creation of an orbital excitation of energy $J$.}
\label{FIG:CIM}
\end{figure}

Up to this point we have studied the bandwidth of holes in an
orbitally disordered state and the instability of the system 
towards orbital-lattice order. In the following we analyze 
to which extent the
itineracy of holes is affected by this disorder-order transition in the 
orbital sector. Namely we are interested in the bandwidth of holes 
moving through an orbitally ordered state; this quantity is then compared 
to our previous result of Eq.\ (\ref{CFL}) for an orbital liquid.
Foremost an important difference between 
models with orbital pseudospin and conventional spin is to be 
noticed here: In the latter systems, spin is conserved when electrons
hop between sites. This implies that hole motion is constrained
in a staggered spin background. In contrast, the transfer Hamiltonian 
(\ref{TRH}) of the orbital model is non-diagonal in orbital pseudospin~---
an orbital basis in which all three transfer matrices 
$t_x^{\alpha\beta}$, $t_y^{\alpha\beta}$, and $t_z^{\alpha\beta}$ 
are of diagonal structure does not exist. This allows holes to move 
coherently even within an antiferro-type orbital arrangement 
[see Fig.\ \ref{FIG:CIM}(a)].
For this reason only a moderate suppression of the hole bandwidth 
is to be expected in the presence of orbital order. We calculate 
the bandwidth for the specific type of orbital order introduced 
in Eq.\ (\ref{UPS}). Starting from the transfer part of Hamiltonian 
(\ref{TRH}) and keeping only the hopping matrix elements that 
allow for a coherent movement of holes (i.e., projecting out
all orbitals which do not comply with the ordered state) we obtain
\begin{equation}
W_{\text{hl}}^{\text{coh}} = 4t.
\label{DCO}
\end{equation}
This result indicates a reduction of the holon bandwidth by
$\approx 30\%$ as compared to the disordered state [see Eq.\ 
(\ref{CFL})]. While not being dramatic, a quenching of the
bandwidth by one third should nevertheless be sufficient to induce 
the localization process, e.g., via the formation of lattice
polarons. However, it is important to note that 
Eq.\ (\ref{DCO}) accounts solely for the coherent motion of holes;
incoherent processes involving the creation and subsequent absorption 
of orbital excitations are neglected 
[see Fig.\ \ref{FIG:CIM}(b)].
Only if the ordered state is robust, i.e., if it costs a large 
amount of energy for an electron to occupy an orbital that does 
not comply with the long-range orbital alignment, these 
incoherent processes become negligible. This limit
does not apply to manganites where the orbital excitation energy
is $J<t$ only. 
Before a conclusion about the role of orbital order in the 
metal-insulator transition can be drawn, these incoherent processes 
have to be investigated in more detail.

In the following we study the influence of incoherent processes onto 
the motion of holes, employing an ``orbital wave'' approximation:
Starting from the assumption that long-range orbital order
has developed and that fluctuations around this ordered state
are weak, we use a slave-fermion representation of the
electron operators in the transfer Hamiltonian (\ref{TRH}):
\[
c^{\dagger}_{i\alpha} = b^{\dagger}_{i\alpha} f_i.
\]
Within this picture, the orbital pseudospin is assigned to bosonic 
orbitons and charge to fermionic holons. The lattice is then divided 
into two sublattices which are ascribed different preferred pseudospin 
directions [see Eq.\ (\ref{UPS})]:
\begin{eqnarray*}
\uparrow &\equiv & \left(|3z^2-r^2\rangle+|x^2-y^2\rangle\right)/\sqrt{2}
\quad \text{on sublattice $A$},\\*
\downarrow &\equiv & \left(|3z^2-r^2\rangle-|x^2-y^2\rangle\right)/\sqrt{2}
\quad \text{on sublattice $B$}.
\end{eqnarray*}
In analogy to conventional spin-wave theory,
excitations around this ground state can be treated by employing the 
following mapping of orbiton operators $b_{i\alpha}$ onto 
``orbital-wave'' operators $\beta_i$:
\[
b_{i\uparrow} = \left\{
\begin{array}{cc}
1 & \text{sublattice $A$},\\
\beta_i & \text{sublattice $B$},
\end{array}\right. \quad
b_{i\downarrow} = \left\{
\begin{array}{cc}
\beta_i & \text{sublattice $A$},\\
1 & \text{sublattice $B$}.
\end{array}\right.
\]
In the momentum representation the transfer Hamiltonian (\ref{TRH}) 
then becomes
\begin{equation}
H_t = \sum_{\bbox{k}} \omega_{\bbox{k}} 
f_{\bbox{k}}^{\dagger} f_{\bbox{k}}+
\sum_{\bbox{k}\bbox{p}} 
\left[ \gamma_{\bbox{k}} \beta^{\dagger}_{\bbox{p}}
+ \gamma_{\bbox{k}+\bbox{p}} \beta_{-\bbox{p}} \right]
f_{\bbox{k}}^{\dagger} f_{\bbox{k}+\bbox{p}}.
\label{SWH}
\end{equation}
Here $\omega_{\bbox{k}} = -t(c_x+c_y-2 c_z)/2$ and 
$\gamma_{\bbox{k}} = t[(2-\sqrt{3})c_x + (2+\sqrt{3})c_y
+2c_z]/2$ with $c_{\gamma} = \cos k_{\gamma}$. The first 
term in Eq.\ (\ref{SWH}) describes the coherent motion of 
holes within a band of width $W_{\text{hl}}^{\text{coh}}=4t$.
The second term describes the interaction of holes
with excitations of the orbital background. The dynamics
of the latter is controlled by the inter-site coupling
Hamiltonian (\ref{HCP}) which in the momentum representation 
becomes
\begin{equation}
H_J = J\sum_{\bbox{k}} \beta^{\dagger}_{\bbox{k}} \beta_{\bbox{k}}.
\label{HOW}
\end{equation}
Hamiltonian (\ref{HOW}) describes dispersionless,
non-propagating orbital excitations of energy $J$.
The local nature of orbital excitations
follows from the absence of frustration effects
in the inter-site orbital coupling term (\ref{HCP}).

The interaction of holes with orbital degrees of freedom 
changes the character of the hole motion: Scattering on orbital
excitations leads to a suppression of the coherent quasiparticle
weight and a simultaneous widening of the holon band.
In analogy to studies of 
spin systems,\cite{KAE89,MAZ91} we analyze these effects 
by employing a self-consistent Born approximation for the
self energy of holes~--- within this method all non-crossing diagrams 
of the self energy are summed up to infinite order while crossing 
diagrams are neglected. Restricting ourselves to the case of a single
hole moving at the bottom of the band, we obtain the following
expression for the holon self energy (see Fig.\ \ref{FIG:SCB}):
\begin{figure}
\noindent
\centering
\epsfxsize=0.5\linewidth
\epsffile{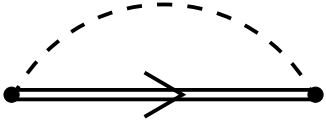}
\caption{Born approximation of the holon self energy:
The double line denotes the ``dressed'' holon propagator 
with self-energy contributions, the dashed line represents 
the ``orbital-wave'' Green's function.}
\label{FIG:SCB}
\end{figure}
\begin{equation}
\Sigma(i\omega) = 
t^2\sum_{\bbox{p}}\gamma^2_{\bbox{p}}\; G(i\omega-J,\bbox{p}).
\label{BRN}
\end{equation}
The Matsubara frequencies are defined as $i\omega = i(2n+1)\pi T$, 
where $T$ denotes temperature and $n$ an integer number.

Our first aim is to study the loss of coherency in the
hole motion. This can be done by employing a dominant-pole
approximation:\cite{KAE89} We split the holon propagator 
in Eq.\ (\ref{BRN}) into its coherent and incoherent parts,
\begin{equation}
G(i\omega,\bbox{k}) = \frac{a_{\bbox{k}}}{i\omega-\tilde{\omega}_{\bbox{k}}}
+G^{\text{inc}}(i\omega,\bbox{k}),
\end{equation}
where $a_{\bbox{k}}$ denotes the quasiparticle weight and
$\tilde{\omega}_{\bbox{k}}$ the not-yet-known renormalized holon 
dispersion. Keeping only the coherent part and using 
$a_{\bbox{k}} = [1-(\partial/\partial \omega) \Sigma'(\omega)]^{-1}$ 
with $\Sigma'(\omega) = \text{Re}[\Sigma(i\omega\rightarrow 
\omega+i0^+)]$, the following recursion relation for the 
quasiparticle weight is obtained:
\begin{equation}
a_{\bar{\bbox{q}}} = \left[1+t^2\sum_{\bbox{p}}\gamma_{\bbox{p}}^2\; 
\frac{a_{\bbox{p}}}{(\tilde{\omega}_{\bar{\bbox{q}}}
-\tilde{\omega}_{\bbox{p}}-J)^2}\right]^{-1},
\label{QPW}
\end{equation}
with $\bar{\bbox{q}}=(0,0,\pi)$ at the bottom of the band.
Equation (\ref{QPW}) can be approximately solved by expanding 
the integrand around $\bar{\bbox{q}}$, which yields
\begin{equation}
a_{\bar{\bbox{q}}} = \left\{\begin{array}{l}
\displaystyle
1-\frac{1}{\sqrt{2\pi^2}}\;\left(\frac{t}{J}\right)^{1/2}
\quad \text{for $J\gg t$},\\[10pt]
\displaystyle
\sqrt[4]{2\pi^2}\;\left(\frac{J}{t}\right)^{1/4}
\quad \text{for $J\ll t$}.
\end{array}\right.
\end{equation}
In the limit $J/t\rightarrow \infty$, orbitals become static and coherent 
hole motion with $a_{\bar{\bbox{q}}}=1$ is recovered. In the opposite limit 
$J/t\rightarrow 0$, the holon quasiparticle weight is completely lost,
indicating strong scattering of holes on orbital fluctuations.

\begin{figure}
\noindent
\centering
\epsfxsize=\linewidth
\epsffile{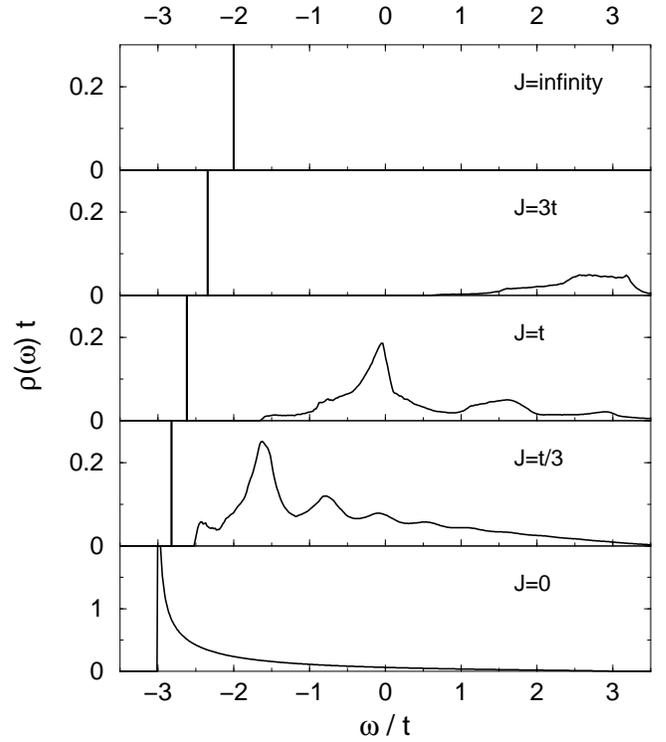}\\[10pt]
\caption{Spectral function of a hole moving at the bottom of the 
band. Different values of $J/t$ are used: The spectrum is 
completely incoherent for $J=0$, having a lower bound at 
$\omega_{\text{min}}=-3t$. With increasing values of $J/t$, spectral weight is 
shifted from the incoherent part of the spectrum to a coherent quasiparticle 
peak (denoted by a vertical line). In the limit $J/t\rightarrow \infty$, the 
quasiparticle peak is at $\omega_{\text{QP}}=-2t$ and has accumulated all 
spectral weight.}
\label{FIG:SPF}
\end{figure}

Next we turn to study the renormalization of the holon bandwidth.
Inserting $G(i\omega,\bbox{k}) = [i\omega-\omega_{\bbox{k}}
-\Sigma(i\omega)]^{-1}$ into Eq.\ (\ref{BRN}) leads
to a recursion relation for the holon self energy:
\begin{equation}
\Sigma(i\omega) = t^2\sum_{\bbox{p}} \frac{\gamma^2_{\bbox{p}}}
{i\omega-\omega_{\bbox{p}}-J-\alpha\Sigma(i\omega-J)}.
\label{BSE}
\end{equation}
The factor $\alpha=(z-1)/z$ partially accounts for the constraint
that forbids more than one orbital excitation per site~---
the hole may therefore not return to a previously visited site 
unless to reabsorb an excitation. We solve Eq.\ (\ref{BSE})
numerically and determine the spectral function
$\rho_{\bar{\bbox{q}}}(\omega) = -\frac{1}{\pi} \text{Im}[G(i\omega\rightarrow
\omega+i0^+,\bar{\bbox{q}})]$ at the bottom of the band. The result is 
shown in Fig.\ \ref{FIG:SPF}.
Different values of $J/t$ are used. In the limit $J/t\rightarrow 0$,
the spectrum is completely incoherent and extends down to 
$\omega_{\text{min}} = -3t$ corresponding to a holon bandwidth of 
$W_{\text{hl}} = 2|\omega_{\text{min}}|=6t$.\cite{REM1} In this limit 
the hole creates its own disorder and effectively moves within 
an orbital-liquid state characterized by strong orbital fluctuations.
In the opposite case $J/t\rightarrow \infty$, all spectral weight
accumulates in a quasiparticle peak (denoted by a vertical line) at
$\omega_{\text{QP}} = -2t$ which corresponds to a bandwidth of 
$W_{\text{hl}} = 2|\omega_{\text{QP}}|=4t$. The orbital state is 
static here and excitations are completely suppressed.
At finite values of $J/t$, the total spectral 
weight is divided into coherent and incoherent parts.
The latter is separated from the quasiparticle
peak by the orbital excitation energy $J$. Processes in which the
hole creates more than one orbital excitation are 
reflected by a succession of peaks in the incoherent spectrum.
For $J=t/3$ which is realistic to manganites, 
the quasiparticle peak accounts for $\approx 65\%$ of the spectral
weight and the width of the holon band is
\begin{equation}
W_{\text{hl}} \approx 5.7 t.
\end{equation}
Comparing the above number with our previous result $W_{\text{hl}} = 6t$ 
for the orbital-liquid state, we find a 
reduction of about $5\%$. We therefore conclude that a disorder-order
crossover in the orbital sector has only a secondary effect
on the kinetic energy of charge carriers, ruling it out as a possible
driving mechanism to initiate the metal-insulator transition
in manganites.

%**********************************************************
%*** Orbital Polarons *************************************
%**********************************************************

\section{Orbital Polarons}

In the preceeding section we have considered charge carriers to 
interact with the orbital sector via the transfer part of 
Hamiltonian (\ref{TRH}). While this coupling
was shown to be responsible for a shift of spectral
weight from the coherent to the incoherent part of the holon
spectrum, the effect onto the full bandwidth was found to
be only small. In the following we point out that in an orbitally
degenerate Mott-Hubbard system there also exists a direct coupling
between holes and orbitals stemming from a polarization of 
$e_g$ orbitals in the neighborhood of a hole. This coupling
is strong enough for holes to form a bound state with the surrounding 
orbitals at low doping concentrations.
Based upon this picture we show that a strong reduction of the 
bandwidth comes into effect as orbital-hole bound states 
begin to form.

%**********************************************************

\subsection{Polarization of Orbitals}

\begin{figure}
\noindent
\centering
\setlength{\unitlength}{0.7\linewidth}
\begin{picture}(1,0.75)
\put(0.355,0.44){\epsfxsize=0.3\unitlength\epsffile{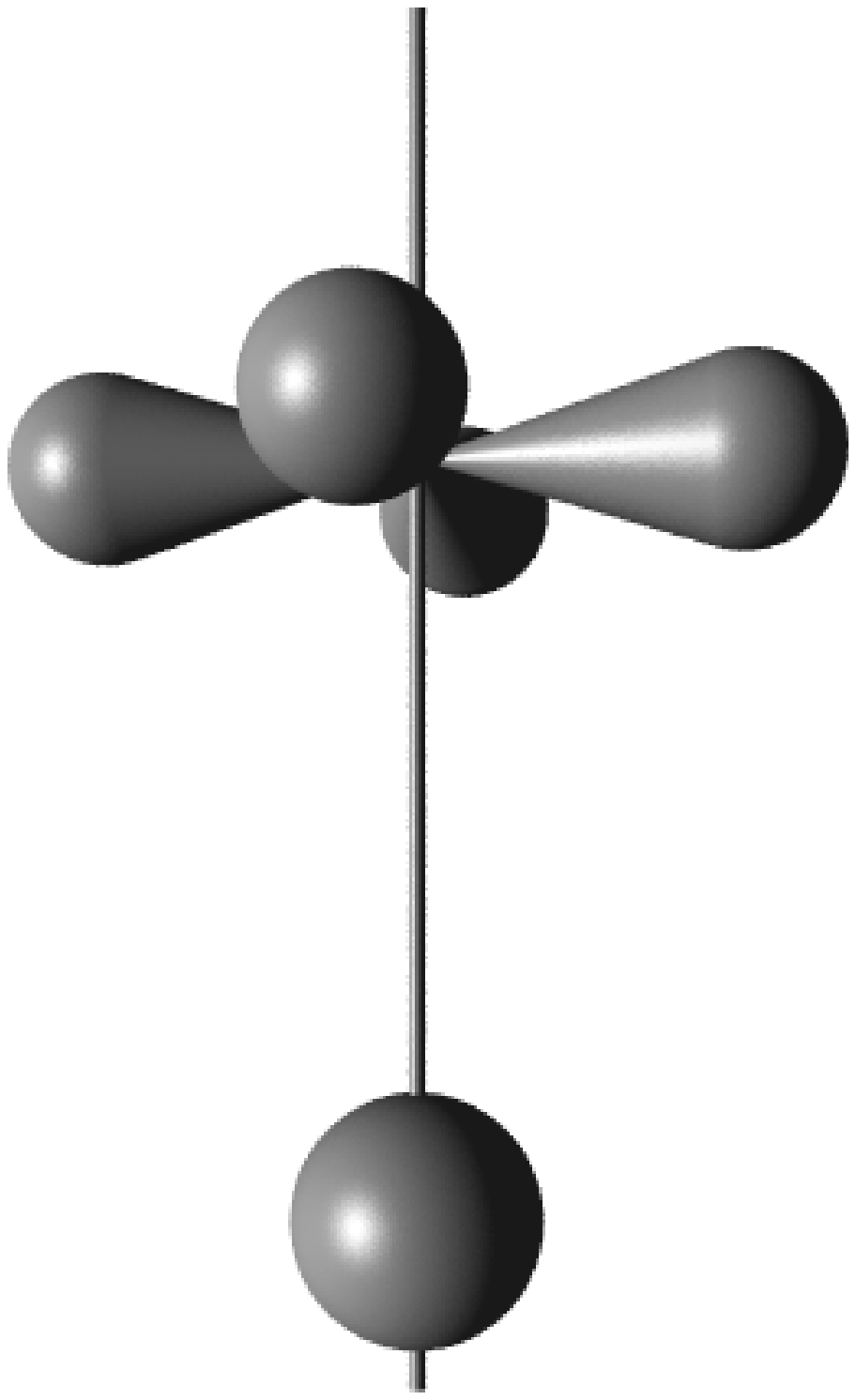}}
\put(0.355,0.01){\epsfxsize=0.3\unitlength\epsffile{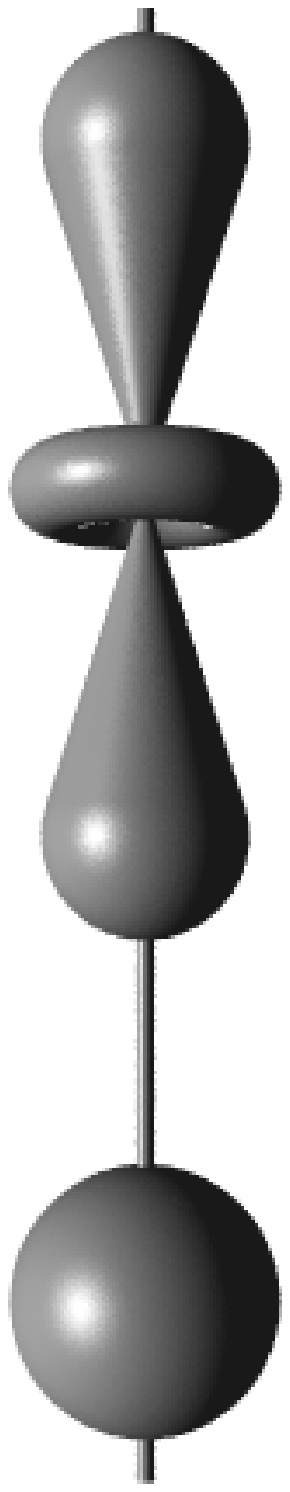}}
\put(0.335,0.44){\fbox{\rule{0.3\unitlength}{0mm}\rule{0mm}{0.3\unitlength}}}
\put(0.335,0.01){\fbox{\rule{0.3\unitlength}{0mm}\rule{0mm}{0.3\unitlength}}}
\put(0,0.15){\epsfxsize=0.85\unitlength\epsffile{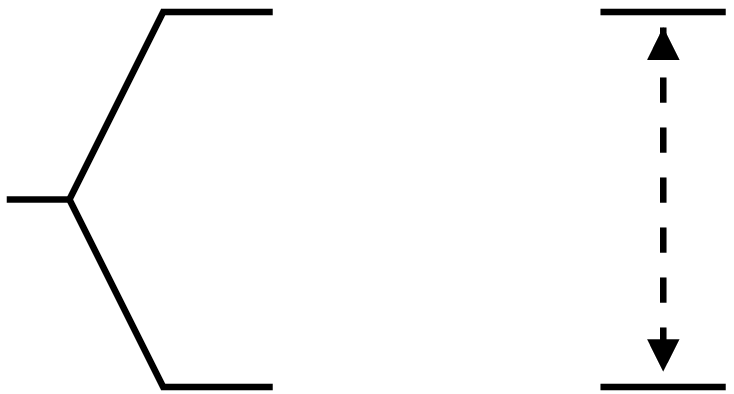}}
\put(0.8,0.36){$\Delta$}
\end{picture}\\[10pt]
\caption{Polarization of $e_g$ levels on sites next to a hole:
Phonons and Coulomb interaction induce a splitting of energy
$\Delta=\Delta^{\text{ph}}+\Delta^{\text{ch}}$.}
\label{FIG:SPL}
\end{figure}

The cubic symmetry of perovskite manganites is locally 
broken in the vicinity of holes which results in a lifting of the
$e_g$ degeneracy on sites adjacent to a hole (see Fig.\ \ref{FIG:SPL}).
Here we discuss two
mechanisms that are foremost responsible for this level splitting: 
(1) a displacement of oxygen ions that move towards the empty site; 
and (2) the Stark splitting of $e_g$ states which is induced by the 
Coulomb force between ``positive'' hole and negative electrons.
The magnitude of the degeneracy lifting 
$\Delta = \Delta^{\text{ph}} + \Delta^{\text{ch}}$ is estimated
as follows: The former phonon contribution $\Delta^{\text{ph}}$ 
originates in the coupling of 
holes to the lattice breathing mode $Q_1$ and of $e_g$ electrons
to two Jahn-Teller modes $Q_2$ and $Q_3$:
\begin{eqnarray}
H_{\text{el-ph}} &=& -\sum_i\Big(
g_1 Q_{1i} n_i^h +g_2 Q_{2i} \sigma_i^x + g_3 Q_{3i} \sigma_i^z
\nonumber\\
&&+\frac{K}{2}\bbox{Q}_i^2\Big),
\label{BJH}
\end{eqnarray}
where $n^h_i$ denotes the number operator for holes and
the Pauli matrices $\sigma^{x/z}_i$ act on the orbital subspace.
The coupling constants are $g_1$ and $g_2\approx g_3$ and 
$K$ is the lattice spring constant. Hamiltonian
(\ref{BJH}) mediates an interaction between empty 
and occupied sites. The effective
Hamiltonian describing this coupling is obtained by integrating 
Eq.\ (\ref{BJH}) over oxygen displacements 
$\bbox{Q}_i=(Q_{1i},Q_{2i},Q_{3i})$.
For a given bond along the $z$ direction this
yields $H^z = -\frac{1}{2}\Delta^{\text{ph}}\;n^h_i\sigma^z_j$ with
\begin{equation}
\Delta^{\text{ph}} = g_1 g_2 \sqrt{2}/(3K) \approx
(g_1/g_2)E_{\text{JT}}.
\end{equation}
A lower bound for this quantity is given by the Jahn-Teller energy,
i.e., $\Delta^{\text{ph}}\ge E_{\text{JT}} \approx 0.2$~eV,\cite{DES98} 
assuming that coupling to the breathing mode is at least as strong as
coupling to the Jahn-Teller modes. Next we estimate the contribution
to the $e_g$-level splitting that follows from the Coulomb interaction 
between a positively charged hole and an $e_g$ electron on a neighboring site.
The magnitude $\Delta^{\text{ch}}$ of this splitting is assessed by taking 
into account the covalence of Mn $3d$ and O $2p$ orbitals, which gives
\begin{equation}
\Delta^{\text{ch}} \approx \frac{3}{4}\gamma^2 R_{\text{Mn}-\text{Mn}}.
\end{equation}
The covalency factor $\gamma = t_{pd}/\Delta_{pd}$ can be
obtained from the transfer amplitude and the charge gap between Mn
and O sites, $t_{pd} \approx 1.8$~eV and $\Delta_{pd}\approx
4.5$~eV.\cite{SAI95} Together with a lattice spacing of
$R_{\text{Mn}-\text{Mn}}=3.9\text{~\AA}$ this leads to 
$\Delta^{\text{ch}}\approx 0.4$~eV. In total the polarization of 
$e_g$ levels on sites next to a hole yields an energy splitting
$\Delta \approx 0.6$~eV. Being comparable in magnitude to the transfer 
amplitude $t$ this number strongly indicates a direct coupling of charge 
and orbital degrees of freedom to be of importance in manganites.

\begin{figure}
\noindent
\centering
\epsfxsize=0.6\linewidth
\epsffile{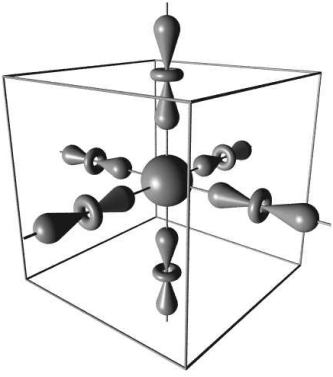}\\[6pt]
\caption{Orbital polaron in the strong-coupling limit: Six $e_g$ 
states point towards a central hole.}
\label{FIG:OPO}
\end{figure}

The splitting of $e_g$ levels effects all six sites surrounding a hole. 
From the above Hamiltonian for a bond along the $z$ axis, analogous 
expressions for $x$ and $y$ directions are derived by a rotation in orbital 
isospin space. The complete Hamiltonian for the cubic system is then
\begin{equation}
H_{\text{ch-orb}} = -\Delta \sum_{\langle ij \rangle_{\gamma}}
n_i^h \tau_j^{\gamma},
\label{HPO}
\end{equation}
with orbital pseudospin operators given by Eq.\ (\ref{PSE}).
Hamiltonian (\ref{HPO}) promotes the formation of orbital polarons.
For low enough hole concentrations these consist of a bound state between
a central hole and the surrounding $e_g$ orbitals pointing towards the hole
as is shown in Fig.\ \ref{FIG:OPO}. 
This configuration also yields a large amplitude of virtual excursions of
$e_g$ electrons onto the empty site. Thus, besides minimizing the 
interaction energy of Hamiltonian (\ref{HPO}), it also allows to lower
the kinetic energy. We note that these virtual hopping processes locally
enhance the magnetic moments of core and $e_g$ spins via the 
double-exchange mechanism, providing a large effective spin of 
the orbital polaron. This naturally explains the development of 
ferromagnetic clusters experimentally observed at temperatures above 
$T_C$.\cite{TER97}

%**********************************************************

\subsection{Binding Energy}

In conventional lattice-polaron theory, the binding energy
is a function of the coupling constant $g$ and the stiffness of 
the lattice which is controlled by the spring constant $K$:
The energy gain stemming from the
interaction between charge carriers and the lattice competes 
against the deformation energy of the crystal. 
In the case of orbital polarons, the underlying picture is very 
similar. Here the coupling constant is given by the 
orbital-charge interaction energy $\Delta$, while the energy scale 
$W_{\text{orb}} = 6xt+J$ is a measure of the ``stiffness'' of 
the orbital sector. These two quantities are expected to 
determine the binding energy of the orbital polaron:
\begin{equation}
E_b^{\text{orb}} = f(\Delta,W_{\text{orb}}).
\end{equation}
The role of $W_b^{\text{orb}}$ can be illustrated as follows 
(see Fig.\ \ref{FIG:COM}):
\begin{figure}
\noindent
\centering
\setlength{\unitlength}{0.35\linewidth}
\begin{picture}(1,1)
\put(0,0){\epsfxsize=\unitlength\epsffile{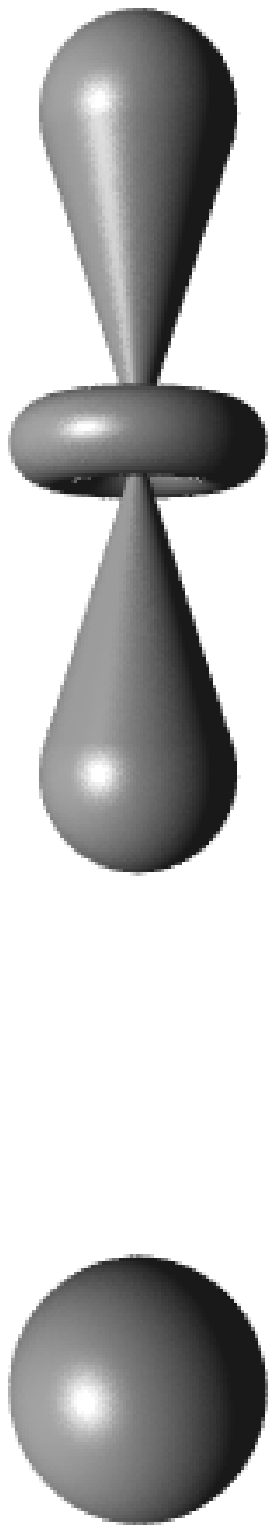}}
\put(0.45,0.242){\epsfxsize=0.095\unitlength\epsffile{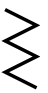}}
\put(0.25,0.8){\epsfxsize=0.5\unitlength\epsffile{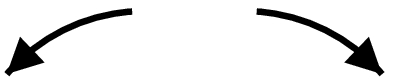}}
\put(0.58,0.30){$\Delta$}
\put(0.75,0.71){$xt$}
\put(-0.1,0.9){(a)}
\end{picture}
\rule{0.3\unitlength}{0mm}
\begin{picture}(1,1)
\put(0,0){
\epsfxsize=\unitlength\epsffile{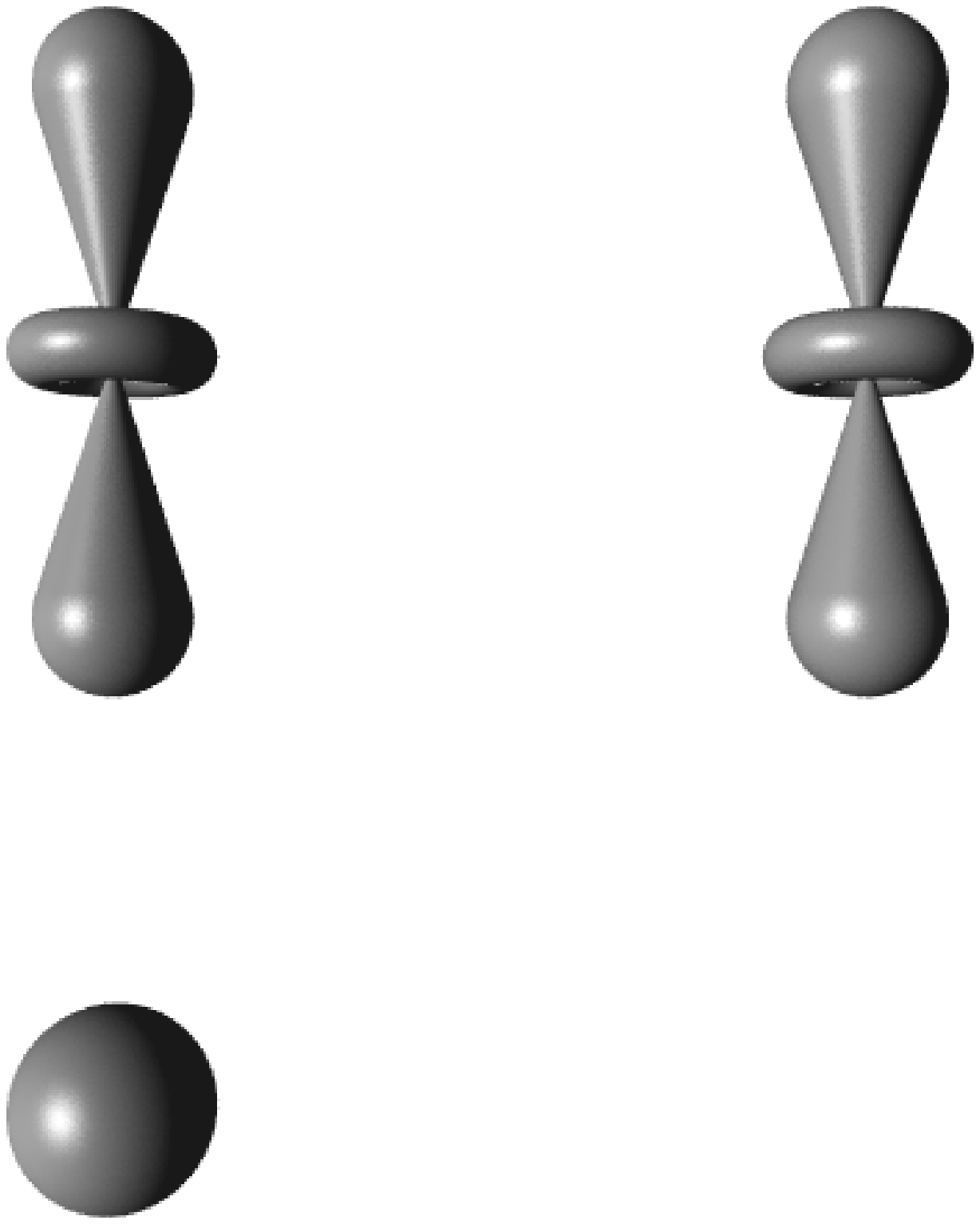}}
\put(0.22,0.242){\epsfxsize=0.095\unitlength\epsffile{zig2.eps}}
\put(0.375,0.63){\epsfysize=0.095\unitlength\epsffile{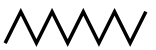}}
\put(0.475,0.755){$J$}
\put(0.35,0.30){$\Delta$}
\put(-0.2,0.9){(b)}
\end{picture}\\[6pt]
\caption{(a) Orbital fluctuations with an energy scale $\propto xt$ and (b)
inter-site correlations $\propto J$ compete against the orbital-hole 
binding energy $\Delta$.} 
\label{FIG:COM}
\end{figure}
In an orbital-liquid state, orbitals have to give up part of their 
fluctuation energy in order to form a bound state with a hole. As a 
consequence, polarons are stable only if orbital fluctuations 
are weak. Furthermore, polarons have a frustrating effect on 
inter-site orbital correlations. The local orientation of orbitals 
favored by Hamiltonian (\ref{HPO}) does in general not comply with 
the orientation that would minimize the Jahn-Teller and superexchange 
coupling between orbitals on nearest- and next-nearest-neighbor sites
of the hole. Thus, in order to form a bound state orbitals have to 
give up part of their inter-site correlation energy $J$ as well. 
The fact that the polaron binding energy is controlled by the 
orbital energy scale $W_{\text{orb}}=6xt+J$ has direct implications 
for the phase diagram of manganites: Due to the doping dependence of
$W_{\text{orb}}$, orbital polarons are stable only at low hole 
concentrations where fluctuations are weak. The tendency of
the system to form polarons is therefore most pronounced in the
lower part of the phase diagram.

\begin{figure}
\noindent
\centering
\epsfxsize=0.9\linewidth
\epsffile{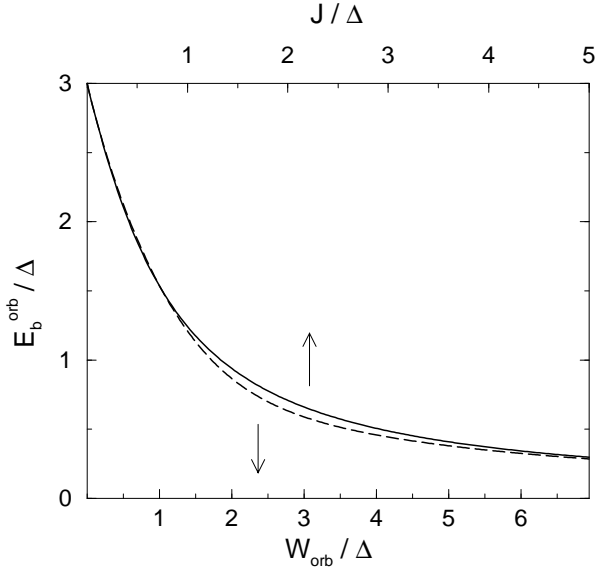}
\caption{Polaron binding energy $E_b^{\text{orb}}$:
The solid curve is a plot of Eq.\ (\protect\ref{EB1}) for the
orbitally ordered state as a function of $J$ (upper axis), the 
dashed curve corresponds to Eq.\ (\protect\ref{EBL}) for the 
orbital-liquid state and is plotted as a function of 
$W_{\text{orb}}$ (lower axis).}
\label{FIG:EOR}
\end{figure}

To derive an expression for the polaron binding energy,
the following approach is used: First we consider a static hole 
placed in an ordered state without fluctuations. We then
calculate the reduction of the total energy due to the interaction 
Hamiltonian (\ref{HPO}). All approximations made in the following 
aim at discarding terms reminiscent of a specific type of orbital 
order, while preserving the most general structure of the orbital-hole 
binding energy. We focus on a single site
located next to the hole in the $z$ direction. The orbital configuration 
at this site is determined by the coupling to the hole described by
Eq.\ (\ref{HPO}) as well as by the orientation of 
neighboring orbitals which couple via superexchange and Jahn-Teller
effect; the latter interaction is determined by Hamiltonian (\ref{HCP}).
Treating all orbitals except the one explicitly considered here
on a mean-field level, the following Hamiltonian is obtained for the
selected site:
\begin{equation}
H^z_{\delta} = -\left(\Delta \tau^z_{\delta} + J \tau^{\Theta}_{\delta}
\right).
\label{HOO}
\end{equation}
Here $\tau^{\Theta}_i = (\sin\Theta \sigma^x_i + \cos\Theta \sigma^z_i)/2$
fixes the orbital orientation which would minimize the interaction
energy with the orbital background~--- for the type of order used in
Eq.\ (\ref{UPS}), e.g., $\Theta=\pm\pi/2$. In general,
this orientation does not coincide with the $|3z^2-r^2\rangle$ configuration 
favored by the orbital polaron which is described by the first term in 
brackets. The state actually chosen by the system
then depends on the ratio of $\Delta$ and $J$ as well as on the angle 
$\Theta$. We determine the energy of this state and, since we are not 
interested in any specific type of order, average over the angle $\Theta$. 
Finally subtracting the $\Delta\rightarrow 0$ limit and 
multiplying with the number of bonds connecting the hole
to its surrounding, the following expression for the polaron binding 
energy is obtained:
\begin{equation}
E_b^{\text{orb}} = 3\left[\sqrt{\Delta^2+J^2}-J\right].
\label{EB1}
\end{equation}
In the limit $J/\Delta \rightarrow \infty$, the orbital state is very
stiff and cannot be polarized by the hole; the binding energy then 
vanishes as $3\Delta^2/(2J)$. On the other hand, the polarization 
is complete in the limit $J/\Delta \rightarrow 0$, yielding a 
maximum value $E_b^{\text{orb}}=3\Delta$ for the binding energy. 
A plot of $E_b^{\text{orb}}$ is shown in Fig.\ \ref{FIG:EOR} (upper curve).
We note that the functional form of Eq.\ (\ref{EB1}) differs from 
the conventional lattice-polaron case where $E_b^{\text{ph}} = 
g^2/(2K)$. This is due to the fact that there exists an
upper limit of the orbital polarization in which the orbitals around
a hole have been fully reoriented to point towards the empty
site (see Fig.\ \ref{FIG:OPO}); technically the existence of this 
upper bound is reflected 
by the hard-core nature of the Pauli operators in Eq.\ (\ref{HOO}).
The familiar form of the binding energy $\propto \Delta^2/J$ 
is recovered only for the weak-coupling limit in which orbital
distortions around the hole are small.

Next we consider a static hole placed in a strongly fluctuating 
orbital-liquid state. Hamiltonian (\ref{HPO}) imposes a splitting 
of $e_g$ levels on the sites next to the hole. The orbiton quasiparticles
of Sec.\ \ref{SEC:ORB} scatter on these local potentials, 
which may lead to the formation of an orbiton-hole bound state.
To calculate the polaron stabilization energy we again consider a 
single site next to the hole in the $z$ 
direction. The local potential imposed by the close-by hole is of 
the form
\begin{equation}
H^z_{\delta} = -\Delta \tau^z_{\delta},
\label{LOP}
\end{equation}
where $\tau^z_i = \frac{1}{2}\sigma^z_i$. We calculate the effect of 
successive scattering of orbitons on the above potential employing
a $T$-matrix formalism; interference between different scattering
centers is neglected here. The correction to the orbiton Green's
function that seizes the effect of Hamiltonian (\ref{LOP}) is given by  
\begin{equation}
\delta G_{\text{orb}}(i\omega;\bbox{R},\bbox{R}) = 
G^0_{\text{orb}}(i\omega;\bbox{R},\delta) T_{\delta}(i\omega)
G^0_{\text{orb}}(i\omega;\delta,\bbox{R}),
\label{DGR}
\end{equation}
with the scattering matrix
\[
T_{\delta}(i\omega) = - \frac{\sigma^z\Delta/2}{1-\sigma^z\Delta
G^0_{\text{orb}}(i\omega)/2}.
\]
Here $G^0_{\text{orb}}(i\omega;\bbox{R},\bbox{R}')$ denotes the orbiton
propagator of the system in the absence of the scattering potential. The 
elements of this 2$\times$2 matrix are given by $[G^0_{\text{orb}}
(i\omega;\bbox{R},\bbox{R}')]^{\alpha\beta} = -\langle T_{\tau} 
f_{\bbox{R}\alpha}f^{\dagger}_{\bbox{R}'\beta}\rangle^0_{i\omega}$ and 
are controlled by the mean-field Hamiltonian (\ref{OHA}). 
The on-site Green's function is $G^0_{\text{orb}}(i\omega) =
\frac{1}{2}\text{Tr}[G^0_{\text{orb}}(i\omega;\delta,\delta)]$.
Integrating over lattice sites, Eq.\ (\ref{DGR}) becomes
\begin{eqnarray}
\delta G(i\omega) &=& - \frac{\Delta}{4}
\left[\frac{\partial}{\partial i\omega}
G^0(i\omega)\right] \nonumber\\*
&&\times\left( \frac{1}{1-\Delta G^0(i\omega)/2}
-\frac{1}{1+\Delta G^0(i\omega)/2}\right),
\label{DGR2}
\end{eqnarray}
with $\delta G_{\text{orb}}(i\omega) = \frac{1}{2}
\sum_{\bbox{R}} \text{Tr}[\delta G_{\text{orb}}(i\omega;
\bbox{R},\bbox{R})]$.
The change in the total energy of the system which is induced by the 
scattering potential can now be obtained from Eq.\ (\ref{DGR2}) by
employing
\begin{equation}
\delta E = 2\int_{-\infty}^{\mu} dw \; (\omega-\mu) \; \delta\rho(\omega),
\label{DEN}
\end{equation}
where $\delta \rho(\omega) = -(1/\pi)\text{Im}[\delta G(i\omega\rightarrow 
\omega+i0^+)]$ denotes the scattering contribution to the density of states.
The orbiton chemical potential is set to $\mu=0$
in the following. We evaluate Eq.\ (\ref{DEN}) by approximating the
on-site Green's function by
\begin{equation}
G^0(i\omega) \approx -\frac{1}{W_{\text{orb}}}
\ln\left[\frac{i\omega-W_{\text{orb}}/2}{i\omega+W_{\text{orb}}/2}\right].
\label{GAP}
\end{equation}
This expression yields a constant density of states for the 
translationally invariant system
which resembles the result that can be obtained numerically 
from the mean-field Hamiltonian (\ref{OHA}).
Approximately solving the integral in Eq.\ (\ref{DEN}) and multiplying 
the result with the number of nearest neighbors of the hole,
we finally arrive at the following expression for the polaron binding
energy:
\begin{equation}
E_b^{\text{orb}} = 3\Delta\left[y\left(\text{cth}\,y-1\right)+
\frac{y\ln 2}{y^2+(\pi/2)^2}\right],
\label{EBL}
\end{equation}
with $y=W_{\text{orb}}/\Delta$. Equation (\ref{EBL}) describes
the polaron stabilization energy in a strongly fluctuating
orbital-liquid state. A plot of this function is shown
in Fig.\ \ref{FIG:EOR} (lower curve). The binding energy reaches
its maximum $E_b^{\text{orb}}=3\Delta$ if orbital fluctuations
are weak, while it vanishes as $3\ln 2 (\Delta^2/W_{\text{orb}})$ 
in the opposite limit where fluctuations are strong.

In the last two paragraphs we have calculated the polaron binding
energy in an orbitally ordered as well as in an orbital-liquid state. 
Although very different approaches were used to describe these
complementary cases, the expressions obtained closely coincide: 
As can be seen in Fig.\ \ref{FIG:EOR},
the two functions are almost indistinguishable if one 
identifies $J$ in Eq.\ (\ref{EB1}) with $\nu W_{\text{orb}}$
in Eq.\ (\ref{EBL}), where $\nu\approx 0.72$ is a numerical fitting
factor. Motivated by the observation that $W_{\text{orb}}=
6xt +J$ reduces to $J$ in the limit of strong orbital correlations 
$J\gg xt$, we discard the fitting factor in the following by 
setting $\nu=1$. We believe these two cases to be smoothly 
connected as should come out in a more elaborate treatment of the 
problem. Based upon these considerations, we conclude that 
Eq.\ (\ref{EB1}) can be 
used to model the polaron binding energy of both the fluctuating 
and the static orbital state:
\begin{equation}
E_b^{\text{orb}} = 3(1-x)\left[\sqrt{\Delta^2+W_{\text{orb}}^2}-
W_{\text{orb}}\right],
\label{EOR}
\end{equation}
where $W_{\text{orb}}=6xt+J$ in former and $W_{\text{orb}}=J$
in the latter case. We finally note that in deriving expressions
(\ref{EB1}) and (\ref{EBL})
for the binding energy, all six sites surrounding the static
hole were assumed to be occupied. Since at finite doping 
the probability of a site being occupied is only $(1-x)$,
we renormalize Eq.\ (\ref{EOR}) by this average occupation
factor.

To summarize, $E_b^{\text{orb}}$ in Eq.\ (\ref{EOR}) represents the
energy to be gained by polarizing the orbital background 
around a static hole. This number depends on the orbital energy
scale $W_{\text{orb}}=6xt+J$, i.e., on orbital fluctuations 
and inter-site orbital correlations which both tend to suppress the 
polaron binding energy. $W_{\text{orb}}$ is to be considered
as the counterpart of the lattice stiffness $K$ in conventional 
polaron theory. The important difference is that $W_{\text{orb}}$
explicitly depends on $x$ which has important consequences for 
the phase diagram of manganites: Orbital polarons can form only at 
low doping concentrations where orbital fluctuations
are weak and the binding energy is consequently large.

%**********************************************************

\subsection{Polaron Bandwidth}

The orbitally degenerate Mott-Hubbard system is instable towards 
the formation of orbital-hole bound states at low doping 
concentrations. In this small-polaron 
regime, holes are pinned by the binding potential
and can move only if being thermally activated. At low
temperatures these processes can be neglected; coherent
charge motion is then possible solely due to quantum tunneling.
Since the polaron is a composite object consisting of a hole 
and several orbitals, the amplitude of these tunneling processes 
is expected to be small as is indeed shown in this section.

\begin{figure}
\noindent
\centering
\epsfxsize=0.9\linewidth
\epsffile{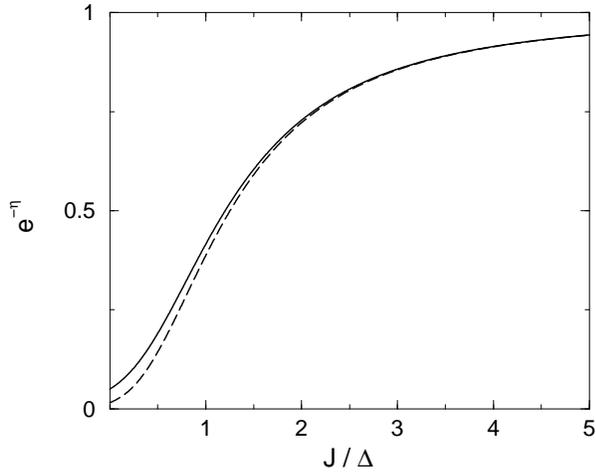}
\caption{Ratio of polaron and holon bandwidth $e^{-\eta}$.
The solid line is based on the exponential
form in Eq.\ (\protect\ref{EET}), the dashed line
represents Eq.\ (\ref{OVL}). The polaron bandwidth shrinks which
decreasing values of $J/\Delta$ associated with an enhanced 
orbital polarizability.}
\label{FIG:ET}
\end{figure}

Here we restrict ourselves to polarons moving in an orbitally 
ordered state. Our analysis is based upon the following idea:
By allowing a hole to polarize the surrounding orbitals, the 
system reduces its ground state energy by $E_b^{\text{orb}}$.
This energy is mostly lost if the hole hops to another site
since the orbital sector cannot immediately adopt to the new 
position~--- orbital fluctuations are slow
compared to those of holes. After a short time
the system returns to the ground state, 
most likely by transferring the hole back to its original location.
But there is also a small probability for the hole
to keep its new position while the orbital sector adapts to the
relocation. This is possible due to the non-orthogonality of 
configurations in which orbitals point towards the old and the
new location of the hole, respectively. The polaron tunneling 
amplitude is then given by the transfer amplitude $t$ of holes
multiplied by the overlap between states with 
orbitals pointing towards the old and new position of the hole, 
respectively. This overlap is calculated as follows: We use 
Hamiltonian (\ref{HOO}) to determine the orientation of a
single orbital next to the hole~--- all other orbitals are 
treated on a mean-field level. As the hole hops, this orbital 
has to change its orientation from pointing towards the hole to 
being aligned with the background. The projection between these 
two states is
\begin{equation}
p = \frac{1}{\sqrt{2}}\left[1+\frac{J}{\sqrt{\Delta^2+J^2}}
\right]^{1/2}.
\end{equation}
An average over the angle $\Theta$ specifying the type of orbital 
order in Eq.\ (\ref{HOO}) has been performed here.
Other orbitals undergo the reversed process: Originally
being aligned with the background, they turn towards the
hole as the latter hops onto a neighboring site.
In total there are $2(z-1)\approx 2z$ orbitals that have to 
reorient. The overlap between the initial and the final
state is then given by $P=p^{2z}$, yielding
\begin{equation}
P = \frac{1}{2^z}
\left[1+\frac{J}{\sqrt{\Delta^2+J^2}}\right]^z.
\label{OVL}
\end{equation}
We rewrite Eq.\ (\ref{OVL}) as
\begin{eqnarray}
P &=& \left[1-\frac{\sqrt{\Delta^2+J^2}-J}{2\sqrt{\Delta^2
+J^2}}\right]^z \nonumber\\
&\approx &\exp\left[-3 \frac{\sqrt{\Delta^2+J^2}-J}
{\sqrt{\Delta^2+J^2}}\right],
\end{eqnarray}
where the exponential form becomes exact for 
large coordination number which has been set to $z=6$.
The denominator in the exponent is identified as the orbital 
binding energy given by Eq.\ (\ref{EB1}). Hence we finally 
arrive at the following compact expression for $P\equiv e^{-\eta}$:
\begin{equation}
e^{-\eta} = \exp\left[-\frac{E_b^{\text{orb}}}{\sqrt{\Delta^2+J^2}}
\right].
\label{EET}
\end{equation}
A plot of this function is shown in Fig.\ \ref{FIG:ET}.
The physical significance of $e^{-\eta}$ is that it relates the
holon to the polaron bandwidth:
\begin{equation}
W_{\text{pol}} = W_{\text{hl}} \; e^{-\eta}.
\end{equation}
As the system becomes critical towards the formation
of orbital-hole bound states, polarons replace holes as
charge carriers. Our result shows that this transition
is accompanied by an exponential suppression of the 
bandwidth. Strictly speaking the translationally 
invariant system remains a metal; in reality, however, the small 
bandwidth makes polarons susceptible to localization, e.g., 
by trapping in the random potential of impurities.
The suppression of the bandwidth is most pronounced
if the polaron binding energy is large: The orbitals
around the hole are then strongly distorted, which 
necessitates a significant reorientation to allow
the hole to hop. We note that the expression in
Eq.\ (\ref{EET}) is similar to the
result obtained in conventional lattice-polaron theory
where $e^{-\eta} = \exp\left(-E_b^{\text{ph}}/\omega_0\right)$.
Here $E_b^{\text{ph}}=g^2/(2K)$ denotes the polaron 
binding energy and $\omega_0$ the phonon frequency; the latter
corresponds to $(\Delta^2+J^2)^{1/2}$ in our orbital-polaron
theory. 

Equation (\ref{EET}) has been derived for a static,
non-fluctuating orbital state. Following the discussion
in deriving the polaron binding energy, we generalize the
result to account for orbital fluctuations as well. This is
done by replacing the inter-site correlation energy $J$ by the 
more general orbital energy scale $W_{\text{orb}}=6xt+J$.
Hence we obtain 
\begin{equation}
e^{-\eta} = \exp\left[-\frac{E_b^{\text{orb}}}
{\sqrt{\Delta^2+W_{\text{orb}}^2}}\right],
\end{equation}
where the polaron binding energy $E_b^{\text{orb}}$ is
now given by Eq.\ (\ref{EOR}).

To summarize, the development of orbital polarons leads to a 
sharp reduction of the bandwidth. In this regime 
the orbital-hole bound state can move only as an entity via 
quantum-tunneling processes. Since the polaron extends over 
several lattice sites, the transfer amplitude is exponentially 
small. The bandwidth reduction is controlled by the ratio 
$E_b^{\text{orb}}/[\Delta^2+W_{\text{orb}}^2]^{1/2}$ which is 
a measure of the orbital distortions around a hole. Strong 
orbital fluctuations and inter-site orbital correlations weaken
the polaron effect by suppressing these distortions.

%**********************************************************
%*** Metal-Insulator Transition ***************************
%**********************************************************

\section{Metal-Insulator Transition}

As was shown in the preceeding section, the formation of 
orbital-hole bound states leads to an exponential suppression 
of the bandwidth which makes the system prone to localization. 
In a double-exchange system, this crossover from a free-carrier 
to a small-polaron picture can be initiated either by a reduction 
of the doping concentration or by an increase in temperature; the
former acts via an enhancement of the polaron binding energy, the 
latter by constricting the motion of holes via the
double-exchange mechanism. In this section we combine our 
orbital-polaron picture with the theory of conventional lattice 
polarons to develop a scheme of the metal-insulator transition in 
manganites.

\begin{figure}
\noindent
\centering
\setlength{\unitlength}{0.9\linewidth}
\begin{picture}(1,0.84)
\put(-0.06,0){
\epsfxsize=\unitlength
\epsffile{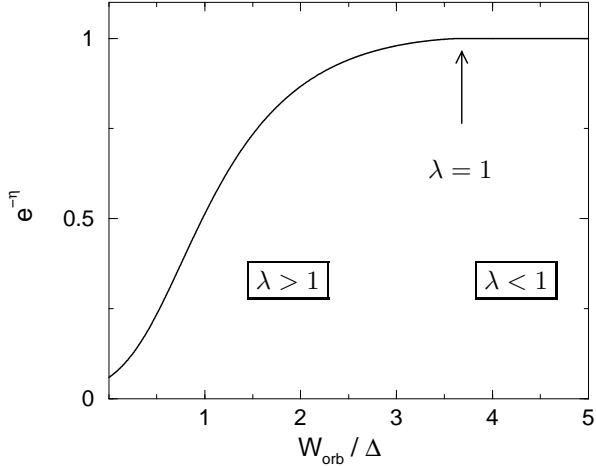}}
\put(0.67,0.51){$\lambda=1$}
\put(0.36,0.32){\fbox{$\lambda>1$}}
\put(0.75,0.32){\fbox{$\lambda<1$}}
\end{picture}
\caption{Crossover from the free-carrier to the small-polarons
regime. At $\lambda<1$ polarons are unstable and the bare holon
bandwidth is recovered. At $\lambda>1$ orbital polarons form, 
leading to an exponential suppression of the bandwidth.
To avoid an abrupt drop of the bandwidth at $\lambda=1$,
Eq.\ (\protect\ref{GAF}) is used to connect the two regimes.}
\label{FIG:GA}
\end{figure}

The transition from a free-carrier to a small-polaron picture is 
governed by the dimensionless coupling constant $\lambda_{\text{orb}} =
E_b^{\text{orb}}/D_{\text{hl}}$, where $E_b^{\text{orb}}$
is the polaron binding energy given by Eq.\ (\ref{EOR})
and $D_{\text{hl}}=W_{\text{hl}}/2=3t$ is the half-bandwidth of 
holes. The lattice breathing mode of Eq.\ (\ref{BJH}) adds an 
additional contribution $\lambda_{\text{ph}} = E_b^{\text{ph}}/
D_{\text{hl}}$ with $E_b^{\text{ph}} = g_1^2/(2K)$, which further 
promotes the formation of polarons. The coupling constant thus 
becomes
\begin{equation}
\lambda = \frac{E_b^{\text{orb}}+E_b^{\text{ph}}}{D_{\text{hl}}}.
\end{equation}
The critical value that separates free-carrier and small-polaron
regimes is $\lambda = 1$.
For $\lambda \gg 1$ small polarons have fully developed and the 
bandwidth is reduced by an exponential factor
\begin{equation}
e^{-\eta} = \exp\left[-\gamma\left(\frac{E_b^{\text{orb}}}
{\sqrt{\Delta^2+W_{\text{orb}}^2}} + 
\frac{E_b^{\text{ph}}}{\omega_0}\right)\right],
\label{ESU}
\end{equation}
with $\gamma=1$. We note that interference effects between
orbital and lattice coupling are neglected in Eq.\ (\ref{ESU}). 
For $\lambda<1$ the free-carrier picture
is recovered and holes move in a band of width 
$W_{\text{hl}} = 6t$. This implies $\gamma=0$ in Eq.\ (\ref{ESU}),
yielding $e^{-\eta} = 1$. To simulate the crossover between
the two regimes, we phenomenologically employ the function
\begin{equation}
\gamma = \left\{
\begin{array}{l}
\displaystyle
\beta-\frac{\ln[\lambda(1+\beta)]}{\lambda^2} \quad \text{for} \quad
\lambda > 1,\\[12pt]
\displaystyle
0 \quad \text{for} \quad \lambda < 1,
\end{array}\right.
\label{GAF}
\end{equation}
with $\beta = [1-1/\lambda^2]^{1/2}$. This function has been
proposed for strongly coupled electron-phonon systems 
(see, e.g., Ref.\ \onlinecite{ALE94}) and avoids an unphysical 
sudden drop of the bandwidth as $\lambda=1$
is reached. The crossover hence obtained for $e^{-\eta}$ 
is depicted in Fig.\ \ref{FIG:GA}.

Up to this point we have mainly focused on the role of the
orbital energy scale $W_{\text{orb}}$ in the formation of small
polarons. We now turn to analyze in more detail the effect of 
temperature. The latter controls the bandwidth  $W_{\text{hl}}$ 
via the double-exchange mechanism: At low temperatures all spins
are ferromagnetically aligned and the transfer amplitude
reaches its maximum. With increasing temperature the
ferromagnetic moment weakens, constricting the motion of
charge carriers. Specifically, the transfer amplitude 
changes with the normalized magnetization $m$ as
\begin{equation}
t = t_0 \sqrt{(1+m^2)/2}.
\end{equation}
The magnetization depends on temperature via the self-consistent 
equation 
\begin{equation}
m = B_S(\alpha m) \quad \text{with} \quad
\alpha = \frac{3S}{1+S} \; \frac{T_C}{T},
\label{MAG}
\end{equation}
where
\[
B_S(y) = \frac{2S+1}{2S} \;\text{cth}\left[\frac{2S+1}{2S}\;y\right]-
\frac{1}{2S}\;\text{cth}\left[\frac{1}{2S}\;y\right]
\]
denotes the Brillouin function. The average magnetic 
moment per site varies with doping: The moments $S^c=3/2$ and $s=1/2$ 
of core and $e_g$ spins combine on average as 
\begin{equation}
S = \frac{3}{2} + \frac{1}{2}(1-x).
\end{equation}
Finally, the Curie temperature $T_C$ in Eq.\ (\ref{MAG}) is controlled by
the strength $J_{\text{eff}}$ of ferromagnetic exchange bonds via
\begin{equation}
T_C = \frac{\nu z}{3}S(S+1) J_{\text{eff}}.
\end{equation}
The fitting parameter $\nu$ compensates for an overestimation of $T_C$ 
in the mean-field treatment. Double-exchange as well as superexchange 
processes are responsible for establishing the ferromagnetic links
between sites. The magnitude of this coupling in the limit of
large Hund's coupling is\cite{KHA99}
\begin{equation}
J_{\text{eff}} = \frac{1}{2S^2} \left[
x(1-x) \chi t e^{-\eta} + (1-x)^2 \frac{2\chi^2 t_0^2}{U}\right].
\label{JEC}
\end{equation}
The first term in squared brackets of Eq.\ (\ref{JEC}) stems from the 
coherent motion of holes/polarons and represents the conventional 
double-exchange contribution to $T_C$. The factor $e^{-\eta}$ accounts 
for the rescaling of the coherent bandwidth
as the small-polaron regime is entered. The second term is due to 
superexchange processes. It describes the high-energy virtual 
hopping of $e_g$ electrons which is insensitive to a 
polaronic reduction of the 
bandwidth. It is also noticed that superexchange in an orbitally
degenerate system is of ferromagnetic nature because of the
large Hund's coupling present in manganites.\cite{KHA99,END99,MAE98}
Superexchange hence dominates the ferromagnetic interaction 
in the small-polaron regime.

The system of equations presented above controls the electronic and
magnetic behavior of manganites at low and intermediate doping
levels. A critical coupling $\lambda=1$ leading to the formation of
polarons can be reached either by lowering the doping concentration
or by increasing the temperature~--- the former enhances $E_b^{\text{orb}}$,
the latter quenches $W_{\text{hl}}$. The equations are interrelated and
have to be solved recursively. As a result of this
self consistency, the breakdown of the metallic bandwidth at
$\lambda=1$ is expected to be rather
sharp: With the evolution of small polarons, the coherent band
width shrinks, thereby weakening the magnetic exchange links. 
Double exchange then drives the system even farther towards the 
strong-coupling limit.

%**********************************************************
%*** Comparison with Experiment ***************************
%**********************************************************

\section{Comparison with Experiment}

To illustrate the interplay between the system of equations
presented in the preceeding section, we numerically extract 
from them the $T$-$x$ phase diagram. While it is obvious that 
$T=T_C$ is a suitable criterion to separate the low-temperature
ferromagnetic from the high-temperature paramagnetic state,
more care has to be taken to distinguish between metallic and
insulating behavior. Our theory describes the reduction of the 
bandwidth which follows from the formation of small polarons. However, 
strictly speaking, the system remains metallic even in the
strong-coupling limit since polarons can still move by tunneling. 
It is therefore necessary to define a critical value of the bandwidth 
beyond which additional effects such as pinning to impurities are 
implicitly assumed to set in and finally turn the system into an 
insulator. The specific criterion used here is only of marginal
importance, as feedback effects discussed above induce a quick collapse 
of the bandwidth once a critical coupling $\lambda=1$ is reached. For 
simplicity we define $\lambda<1$ to be a metal and $\lambda>1$ to be 
an insulator.

\begin{figure}
\noindent
\centering
\epsfxsize=0.92\linewidth
\epsffile{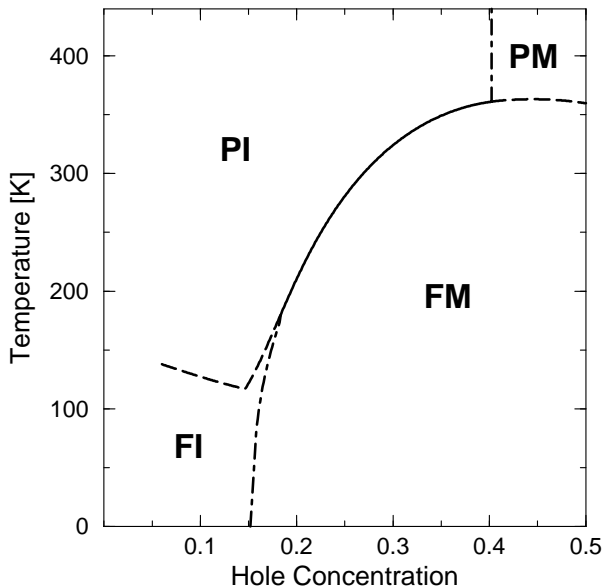}
\caption{Magnetic and electronic phase diagram obtained within
the present theory. Dashed and dot-dashed lines represent 
magnetic and electronic transitions, respectively, the simultaneous 
transition in both channels is denoted by a solid line. The phases are: 
paramagnetic insulator (PI), paramagnetic metal (PM), ferromagnetic 
metal (FM), and ferromagnetic insulator (FI).}
\label{FIG:PHD}
\end{figure}

The following parameters are chosen for calculating the phase diagram:
The orbital polarization energy is set to 
$\Delta = 0.55~\text{eV}$, yielding a binding energy comparable to 
the phononic one $E_b^{\text{ph}} = 0.45~\text{eV}$; the phonon frequency 
is $\omega_0 = 0.05~\text{eV}$, the interaction between orbitals 
$J = 0.13~\text{eV}$, the bare transfer amplitude 
$t_0 = 0.36~\text{eV}$,\cite{KIL98} and $U = 4.0$~eV. The fitting 
parameter $\nu=0.55$ (Ref.\ \onlinecite{REM2}) is 
adjusted to reproduce the values of $T_C$ 
observed for La$_{1-x}$Sr$_x$MnO$_3$.\cite{URU95} 
The result is shown in Fig.\ \ref{FIG:PHD}.

Our most important observation is that the doping dependence of 
orbital polarons makes the system more insulating at low and more 
metallic at high doping levels. Convincingly this is seen in the complete 
absence of metalicity at $x<0.15$ and the appearance of a metallic 
phase above $T_C$ at $x>0.4$. 
The region $0.15 < x < 0.4$ in which colossal magnetoresistance is
experimentally observed is characterized by a simultaneous 
magnetic and electronic transition from a ferromagnetic metal to a 
paramagnetic insulator.\cite{SCH95} The role of polarons in this transition
is most pronounced at low hole concentrations. This can be seen
from the behavior of the magnetization as $T_C$ is approached from
below (see Fig.\ \ref{FIG:MAG}):
\begin{figure}
\noindent
\centering
\epsfxsize=0.9\linewidth
\epsffile{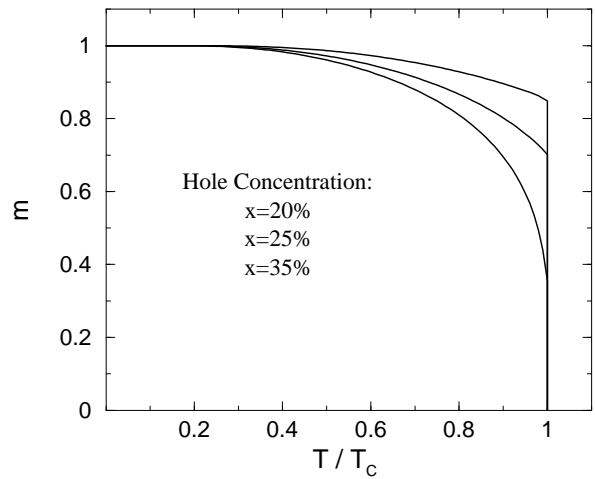}
\caption{Variation of the normalized magnetization $m$ with temperature.
At low hole concentrations (upper curve), the system is close
to a transition to the small-polaron regime~--- a small increase
in temperature then already sufficiently impedes the hole motion for
polarons to form, resulting in a sharp drop of the magnetization.
At large $x$ (lower curve), the conventional double-exchange picture 
is recovered.}
\label{FIG:MAG}
\end{figure}
At low $x$, the polaron binding energy is large and the system is close 
to localization. A small reduction of the bandwidth via double exchange
is then sufficient to trigger the formation of polarons, resulting in a 
sudden collapse of the magnetic moment. Such a sharp drop 
signals the presence of a localization mechanism beyond
double exchange and is indeed seen experimentally
(see, e.g., Refs.\ \onlinecite{SCH95,ZHA96,FRA99}). 
On the other hand, at larger hole concentrations the polaron binding 
energy is comparably small. Thus, a significant suppression of
the bandwidth via double exchange is needed before polaron formation 
can set in. The magnetization curve now closely resembles the one 
predicted by double-exchange theory.

Clearly beyond the grasp of conventional double-exchange theory
lies the emergence of ferromagnetism in the insulating phase
at low doping. Mostly responsible for this are superexchange processes 
which mediate a ferromagnetic interaction even in the insulating phase.
Ferromagnetism is further promoted by the existence of 
orbital polarons: Charge fluctuations 
inside the polaron provide a strong local ferromagnetic coupling between 
sites close to a hole, hence establishing ferromagnetic 
clusters seen in experiment.\cite{TER97}
At sufficiently large hole densities these clusters start to interact,
thereby forming a ferromagnetic state.

As was discussed above, orbital fluctuations are predominantly induced by 
the motion of holes. The loss of charge mobility in the insulating phase 
should therefore trigger static orbital order. An
orbitally ordered state has in fact been experimentally detected in the 
insulating regions of La$_{0.88}$Sr$_{0.12}$MnO$_3$.\cite{END99} However, 
in general such an ordered state is expected to have orbital and 
Jahn-Teller glass features due to the presence of quenched orbital 
polarons, thereby reducing the uniform component of Jahn-Teller distortions. 
Finally it is worth to notice that the phase diagram in this 
theory is highly sensitive to the transfer amplitude $t_0$ as this parameter 
enters in the polaron binding energy.

%**********************************************************
%*** Conclusion *******************************************
%**********************************************************

\section{Conclusion}

In summary, we have shown that a spontaneous development of 
orbital-lattice order is in general insufficient to trigger the
localization process in manganites. Rather an additional
mechanism was identified: Orbital polarons were illustrated 
to represent an intrinsic feature of an orbitally degenerate 
Mott-Hubbard system and to play an important role in the physics 
of manganites. The binding energy of these orbital-hole bound states 
depends on the rate of orbital fluctuations and hence on the
concentration of doped holes: 
Polarons can form at low doping levels where orbitals fluctuate
only weakly but they become unstable at higher levels of $x$.
This scheme naturally introduces the hole concentration as
an additional variable into the localization process. Most striking
in this respect is the complete breakdown of metalicity 
observed below a critical hole concentration despite the
fact that the system remains ferromagnetically ordered.
On the other hand, orbital polarons become negligible at larger
doping levels where the theory presented here converges onto a 
lattice-polaron double-exchange picture. Accounting
for both orbital and lattice effects we are finally able to reproduce
well the important aspects of the phase diagram of manganites.
In general it can be concluded that a direct coupling between
holes and surrounding orbitals is of crucial importance for the
physics of manganites; its implications extend
clearly beyond the metallic phase alone and can be expected
to play an important role throughout the whole phase diagram.

%************************************************************
%***** References *******************************************
%************************************************************


\begin{references}
\bibitem{REM1} The factor $\alpha=(z-1)/z$
in Eq.\ (\protect\ref{BSE}) compensates only partially for the
mean-field treatment of the constraint that forbids more than
one orbital excitation per site. The bandwidth numerically
obtained for $J/t\rightarrow 0$ is therefore slightly larger
than $W_{\text{hl}} = 6t$.
\end{references}
\end{document}